Jan Oberst

# Sliding Block (Slick) Hashing: An Implementation & Benchmarks

Exercise Submission "Algorithm Engineering"

With hash tables being one of the most used data structures, Lehmann, Sanders and Walzer [1] propose a novel, light-weight hash table, referred to as *Slick Hash*. Their idea is to hit a sweet spot between space consumption and speed. Building on the theoretical ideas by the authors, an implementation and experiments are required to evaluate the practical performance of Slick Hash. This work contributes to fulfilling this requirement by providing a basic implementation of Slick Hash, an analysis of its performance, and an evaluation of the entry deletion, focusing on the impact of backyard cleaning. The findings are discussed, and a conclusion is drawn.

## Methodology

The implementation is done using Rust, making use of its high performance and memory-safety. For benchmarking, the *Oxidized Rust Framework* provided by Gerd Augsburg is run on my local machine. The required functions – namely *insert*, *at* and *contains* – are realized through the *try_insert* and *get* methods, while *contains* simply calls the get method and returns whether the element was found or not. Additionally, the interface exposes a *delete_entry* method which includes backyard cleaning.

The code is publicly available on GitHub [2].

## Analysis

For the analysis, the hyper parameters provided in [1] are used as a starting point. First, the influence of the hyper parameters is studied in a grid search manner. Second, Slick Hash is benchmarked in comparison to the HashMap and the BinaryTreeMap implementations from the Rust standard library. It is worth noting that the performance metrics measurement is implemented in a way that neither memory consumption nor the execution time are affected. All benchmarks are performed with a fixed capacity of 2,000,000 and if not stated otherwise with 2,000,000 insertions.

### Hyper Parameter Evaluation

Both memory consumption and speed of Slick Hash are significantly influenced by the hyper parameters – namely the initial block size $B$, the sliding block size $\hat{B}$, the maximum offset $\hat{o}$, and the maximum threshold $\hat{t}$. To investigate this, an initial configuration of $B = 10, \hat{B} = 2B, \hat{o} = \hat{t} = B$ is defined based on experiments and the proposals by the authors of Slick Hash. To analyze the effects of the modification of the parameters, different values are investigated in a grid search manner while the other parameters remain constant. The sliding block size and the maximum offset are modified jointly because it is only possible to use a larger sliding block size if the maximum offset allows for it. Table *1* displays the hyper parameters used.

Table 1: Hyper parameters used for analysis (default values in blue).

| | | | | |
|---|---|---|---|---|
| $B$ | 5 | 10 | 50 | 200 |
| $\hat{B}$ | 2B | 4B | $B^2$ | - |
| $\hat{o}$ | B | 2B | $\frac{1}{2}B^2$ | - |
| $\hat{t}$ | B | 4B | $B^2$ | - |

Slick Hash is evaluated in terms of its execution time, cache misses, and branch misses for both insertion and querying, and the results are presented in Figure *1*. Three main phenomena can be observed. First, the block size has a major influence on the space efficiency of the hash table. The larger the block size, the less meta data must be stored. Additionally, more values are bumped to the backyard leading to an even greater memory consumption for smaller block sizes (see Appendix for information on the backyard sizes of the different configurations after insertion). Second, with greater block size come greater querying and insertion times. This appears to be quite intuitive since the runtime of both operations depends on the block size.

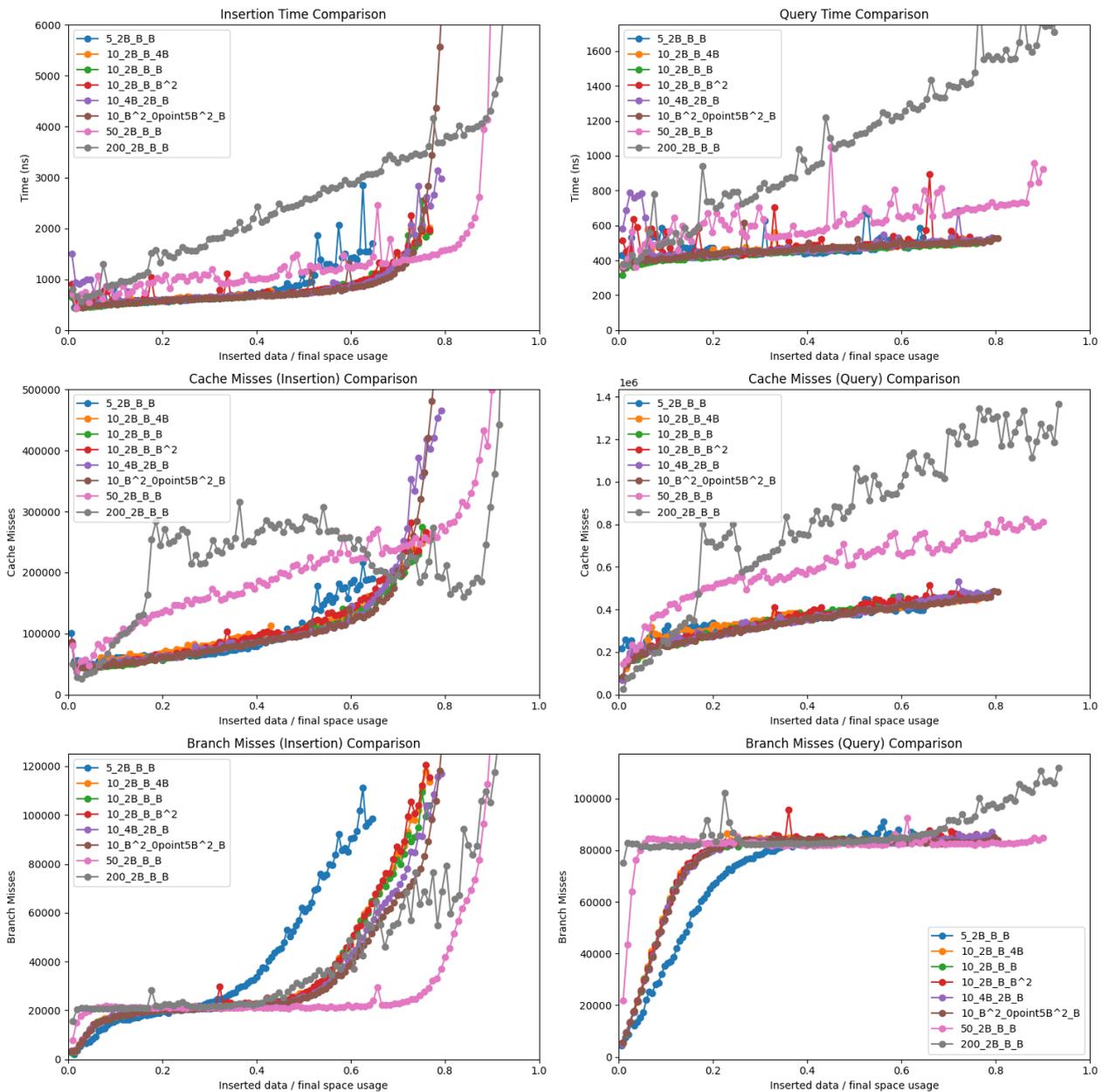

Figure 1: Performance evaluation of Slick Hash hyper parameters: Execution time, cache misses, and branch misses for insertion and querying for different configurations. Legend: BlockSize_SlidingBlockSize_MaximumOffset_MaximumThreshold

Third, tables with larger block sizes tend to expose more cache misses, both during insertion and querying. For the time being, no obvious explanation could be found regarding this phenomenon.

Both the data and the Python script used for plotting are available in the [Git repository](). Additional plots for the comparison of the various Slick Hash configurations with the other two baseline hash tables can be found there.

## Benchmarking

The intuition of Lehmann, Sanders and Walzer about the hyper parameter choice turned out to represent the best configuration in terms of trade-off between memory consumption and speed. The performance of this Slick Hash configuration in comparison to HashMap and BinaryTreeMap is displayed in Figure 2. It can be observed that Slick Hash outperforms the BinaryTreeMap in almost all metrics and can even compete with the HashMap implementation when querying. Moreover, the lightweight structure of Slick Hash is highlighted by the comparably high memory efficiency.

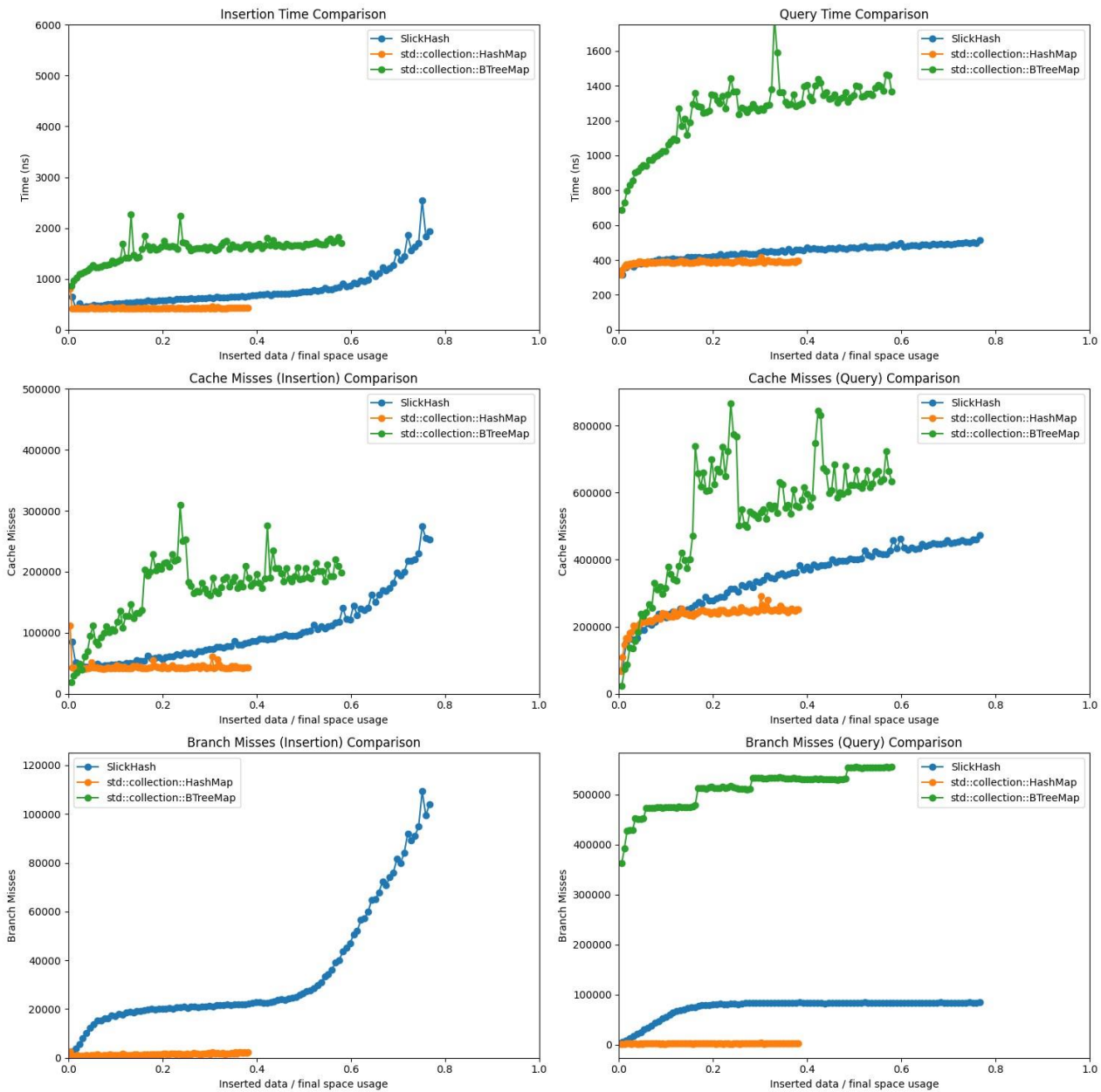

*Figure 2: Benchmarking Slick Hash in comparison to HashMap and BinaryTreeMap implementations from the Rust standard library regarding execution time, cache misses, and branch misses during insertion and querying*

## Entry Deletion

The entry deletion is implemented based on the ideas provided in the Slick Hash paper. To benchmark the implementation, a table is created, and 2,000,000 entries are inserted for Slick Hash, HashMap, and BinaryTreeMap. All entries are deleted subsequently as batches and the average deletion time is measured. The comparison is depicted in Figure *3*.

The cleaning of the backyard was implemented in a naïve way. Every time an entry is deleted, a check is performed whether the main table has enough space to accommodate all entries currently in the backyard. This however leads to unfeasible execution times which is why it was not possible to compare the version with and without the backyard cleaning. Another challenge when cleaning the backyard is that keys with small threshold values are more likely to enter the backyard table. It therefore remains unlikely or even impossible for them to return to the main table if only few or even no entries are inserted in the meantime. This problem could be addressed in future work.

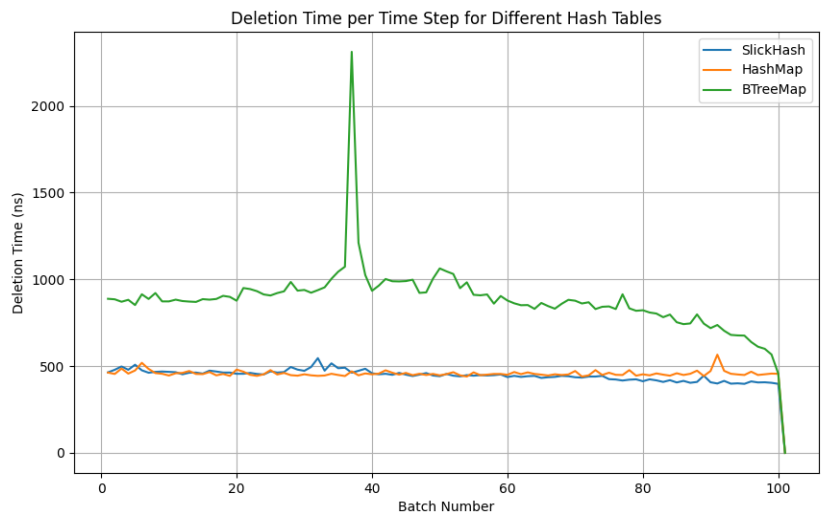

Figure 3: Comparison of deletion time of Slick Hash, HashMap, and BinaryTreeMap

# Discussion

The observations of the evaluations above highlight the viability of Slick Hash and motivate further investigations to push the limits of its performance. Especially regarding the trade-off between memory consumption and speed, Slick Hash seems to have found a sweet spot. On the flip side, there are still weaknesses that need to be improved upon. For example, the static implementation currently suffers from increasing insertion times and slightly increasing querying times for higher load factors. As mentioned earlier, cleaning up the backyard also seems infeasible with the current techniques.

For me, an insight of this work is the challenge associated with moving from a theoretical idea of a data structure to a practical implementation. Although Slick Hash is not of a complex nature per se, practical implementations often uncover problems not foreseen during theoretical analysis. For example, one must be weary of a block growing to the right, over-arcing entire empty blocks, and squishing them in between the subsequent blocks. This happens especially often when dealing with small block sizes. In the presented implementation, this is prevented by not sliding left or right if the block providing a free slot is empty and only has a gap of one. Furthermore, the check if the maximum offset would be exceeded on the right-most block when sliding is missing in the pseudo code provided by [1]. Many details like that are encountered during implementation making it vital to not only focus on theoretical analysis but also implementing ideas in practice.

# Conclusion

In summary, this report explores and evaluates the Sliding Block (Slick) Hashing algorithm, which aims to strike a balance between space efficiency and speed in hash tables. Through a Rust implementation and benchmarking against standard libraries, this study finds that the initial hyper parameters offer an optimal trade-off. Despite challenges in entry deletion, Slick Hash demonstrates promise, highlighting the need for practical implementation considerations. To really compete with existing implementations, further research is required looking into the details of cache efficiency, compiler options and possibilities for parallelism. While weaknesses persist, the algorithm presents a valuable approach, paving the way for future enhancements and emphasizing the importance of bridging theoretical concepts with practical implementation.

# References

[1] Lehmann, Hans-Peter, Peter Sanders, and Stefan Walzer. "Sliding Block Hashing (Slick)--Basic Algorithmic Ideas." arXiv preprint arXiv:2304.09283 (2023).

[2] Oberst, Jan. "SlickHash: Sliding Block Hashing Implementation." Accessed: 2024-09-30. URL: https://github.com/JanIsHacking/slick-hash. (2024).

# Appendix

Table 2: Number of elements in the backyard after insertion of 2,000,000 entries for the different hyper parameter configurations

| Hyper Parameter Configuration (Block Size, Sliding Block Size, Maximum Offset, Maximum Threshold) | Number of Elements in the Backyard | Total Number of Elements in the Table |
|---|---|---|
| 5, 2B, B, B | 153,638 | 2,000,000 |
| 10, 2B, B, B | 78,134 | 2,000,000 |
| 50, 2B, B, B | 16,771 | 2,000,000 |
| 200, 2B, B, B | 4,342 | 2,000,000 |
| 10, 4B, 2B, B | 44,888 | 2,000,000 |
| 10, B^2, (B^2)/2, B | 19,405 | 2,000,000 |
| 10, 2B, B, 4B | 74,252 | 2,000,000 |
| 10, 2B, B, B^2 | 73,671 | 2,000,000 |